\documentclass[]{pasj02} 
\usepackage[switch,mathlines]{lineno} 
\usepackage{url}
\usepackage{fix-cm}

\jyear{2024}
\Received{}
\Accepted{}

\begin{document} 

\title{The Eclipsing $\gamma$ Doradus Star V421 Pegasi}

\author{Jae Woo \textsc{Lee}\altaffilmark{1}\altemailmark\orcid{0000-0002-5739-9804}} 
\email{jwlee@kasi.re.kr}

\altaffiltext{1}{Korea Astronomy and Space Science Institute, Daejeon 34055, Republic of Korea}


{}

\maketitle

\begin{abstract}
We present high-precision TESS photometry of V421 Peg (TIC 301747091), an early F-type eclipsing binary containing a candidate 
$\gamma$ Dor component. The observed short-cadence data allow the detection of pulsation signals, along with revision of 
the fundamental properties of the component stars. Detailed binary modeling indicated that the program target is 
a partially-eclipsing detached system in a circular orbit and that both components are currently in super-synchronous states. 
The radii of each star were measured with an accuracy of about 1 \%. By periodogram analysis of the outside-eclipse residual 
lights obtained from the binary star model, we extracted nine significant signals, five of which are likely aliasing frequencies 
due to sampling artifacts and uncorrected trends in the data used. The other signals of $f_1$, $f_2$, $f_3$, and $f_6$ are 
considered to be independent pulsations with frequencies ranging from 0.73 day$^{-1}$ to 1.02 day$^{-1}$, corresponding to pulsation 
constants of 0.63$-$0.88 days. These frequencies, pulsation constants, and position on the H-R diagram reveal that the pulsating 
signals are $\gamma$ Dor variables arising from the V421 Peg primary component. 
\end{abstract}


\section{Introduction}

Double-lined eclipsing binaries (DEBs) allow direct and accurate measurements of fundamental astrophysical parameters through 
two types of observations such as space-based photometry and high-resolution echelle spectroscopy. In particular, detached DEBs, 
with no mass exchange and little other binary effects, represent typical single stars, and determining their masses and radii to 
better than 2 \% precision is essential for calibrating and improving theoretical stellar models \citep{Andersen1991,Southworth2015}. 
Meanwhile, the component stars of a binary system are in various stages of evolution and can pulsate when they are in instability 
strips on the Hertzsprung-Russell (H-R) diagram. The pulsation features are useful and effective for understanding the star's 
internal structure and physics through asteroseismology. Therefore, oscillating DEBs are of great interest in stellar astronomy 
due to the powerful synergy of these two properties. 

Among pulsating stars, $\delta$ Sct and $\gamma$ Dor variables are intermediate-mass dwarfs of spectral types A-F, with similar 
physical parameters and partially overlapping instability strips, but with significant differences in pulsation frequencies ($f$) 
and constants ($Q$) \citep{Breger2000,Handler+2002,Warner+2003,Uytterhoeven+2011}. Typically, the former pulsates in a frequency 
range higher than $\sim$4 day$^{-1}$ with pulsation constants of $Q <$ 0.04 days, while the latter pulsates below this criterion 
with larger values of $Q >$ 0.23 days. The $\delta$ Sct pulsations are low-order pressure modes triggered by the $\kappa$ 
mechanism \citep{Aerts2010,Balona2015,Antoci+2019} and the $\gamma$ Dor pulsations are high-order gravity modes excited by 
a flux-blocking mechanism \citep{Guzik+2000,Dupret+2004,Dupret+2005}. 

The TESS space mission has been consistently discovered new pulsating EBs, including A-F dwarfs. Most of these exhibit 
$\delta$ Sct pulsations, while only a limited number exhibit $\gamma$ Dor pulsations \citep{Chen+2022,Kahraman+2022,Shi+2022,Southworth+2022}. 
Understanding these two types of pulsations requires accurate fundamental parameters, which can be obtained by analyzing 
time-series light curves combined with the radial velocities (RVs) of both eclipsing components. However, due to the absence 
of good light curves or spectroscopic data, the physical properties of many pulsating EBs remain poorly understood. 
Meanwhile, not all stars within the instability region pulsate, so it would be very interesting to investigate the differences 
between pulsating and non-pulsating stars \citep{Guzik+2015,Kahraman+2025}. 

This study is dedicated to improving understanding of the physical properties of the detached DEB V421 Peg (TIC 301747091, 
HIP 578, ASAS J000702+2250.7, TYC 1729-206-1, Gaia DR3 2847328300933441152; $T_{\rm p}$ = $+$7.936) by discovering and 
characterizing its multi-period pulsations, through detailed analysis of high-precision short-cadence TESS data. 
The target star was announced as an EB with a period of 1.54 days and spectral type F0 from the Hipparcos satellite, and 
has since been the subject of several photometric surveys. Most recently, \citet{Ozdarcan+2016} reviewed the system's 
historical details, and measured the double-lined RVs and atmospheric parameters from their own spectra. Combining 
the spectroscopic measurements with the Hipparcos \citep{ESA1997} and ASAS \citep{pojmanski1997} light curves, they reported 
that the binary star is an early F-type detached EB with masses of $1.594\pm0.029$ $M_\odot$ and $1.356\pm0.029$ $M_\odot$, 
radii of $1.584\pm0.028$ $R_\odot$ and $1.328\pm0.029$ $R_\odot$, and luminosities of $6.25\pm0.0.42$ $L_\odot$ and 
$3.78\pm0.29$ $L_\odot$. The components' positions on the H-R diagram suggests that the primary is a candidate for 
an intermediate-mass pulsator. However, they were unable to detect any intrinsic variability in the observations used, 
probably because the data qualities and observing cadences were not sufficient to investigate the presence of pulsations.   


\section{TESS Photometry and Orbital Ephemeris} 

V421 Peg was observed in two sectors as part of the TESS space mission \citep{Ricker+2015}. The photometric data from sector 57 
(S57) were taken at 120-s cadence, and from sector 84 (S84) at 200-s cadence. \citet{Ozdarcan+2016} reported that the binary star 
is composed of two main-sequence (MS) dwarfs of early-F spectral type, making them candidates for $\gamma$ Dor and/or 
$\delta$ Sct variables. To detect multi-period, high-frequency pulsations, this work concentrated on the 120-s observations 
obtained at higher sampling. The S57 photometry was conducted between 2022 September 30 and October 29 (BJD 2,459,853.35 $-$ 2,459,2882.12) 
and its reduced SPOC data \citep{Jenkins+2016} were obtained via MAST\footnote{https://archive.stsci.edu/}. The CROWDSAP factor 
for our target star is 0.99559623, which means that $\sim$0.4 \% of the TESS flux measurements could come from nearby sources in 
the photometric aperture. 

\begin{figure}
\begin{center}
\includegraphics[scale=0.9]{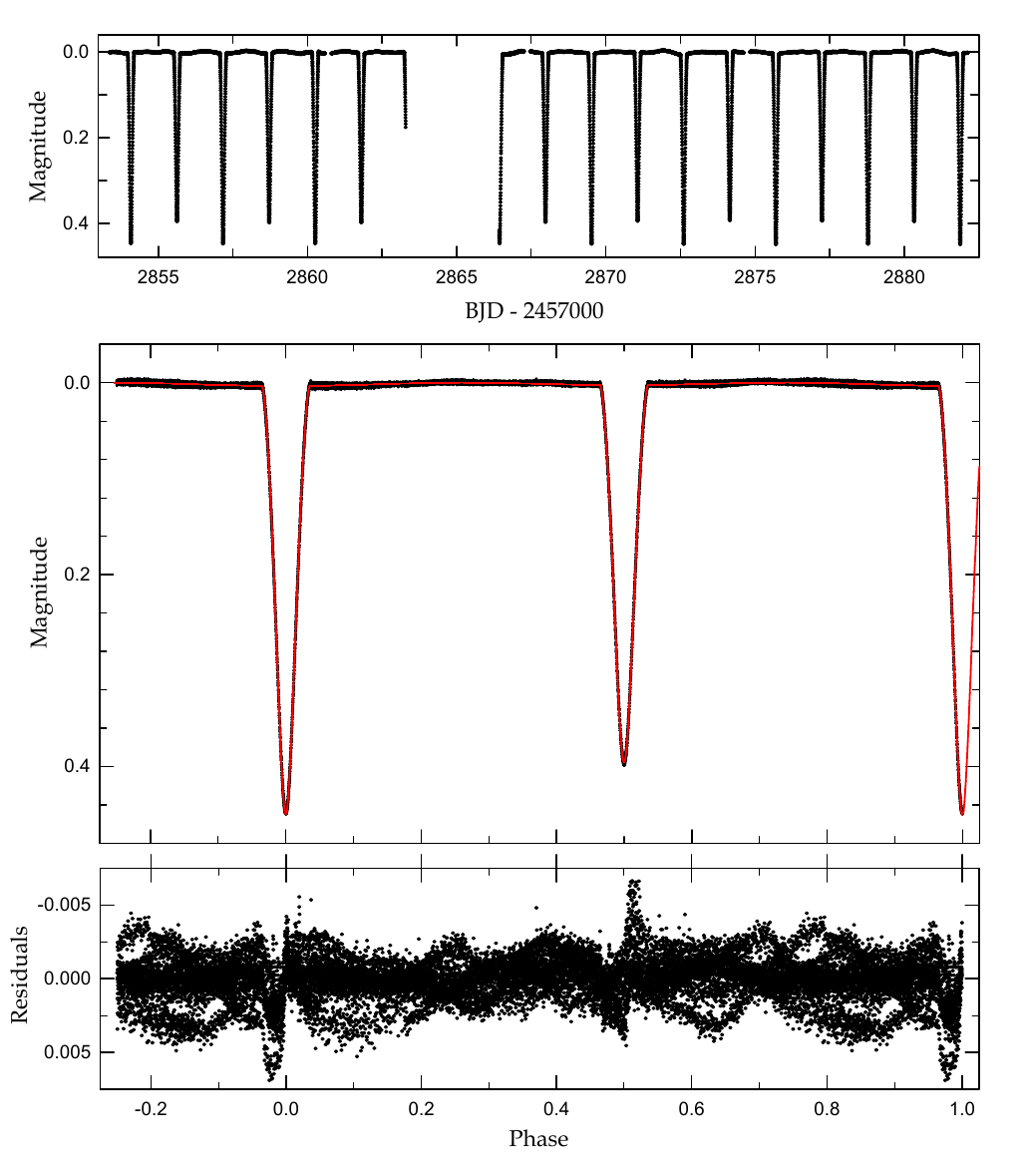}
\end{center}
\caption{TESS observations of V421 Peg distributed in BJD (top panel) and orbital phase (second panel). The circles are 
individual measurements and the solid line represents the synthetic curve obtained with our binary modeling. The corresponding 
residuals are plotted in the bottom panel. \\
{Alt text: TESS light curve observed at 120-s cadence in Sector 57.} }
\label{Fig1}
\end{figure}

We used detrended SAP data with a quadratic fit to the non-eclipsing part \citep{Lee+2024}, which were transformed to a magnitude 
scale and normalized to a maximum light of 0.0 mag. To obtain an orbital ephemeris suitable for the TESS data, we first measured 
the minimum epochs from each eclipse curve \citep{Kwee+1956}, and then applied a linear least-squares fit to them, as follows: 
\begin{equation}
 \mbox{Min I} = \mbox{BJD}~ 2,459,854.082518(71) + 3.087557(12)E,  
\end{equation}
where the parenthesized numbers are the 1$\sigma$ errors for the last two digits of each coefficient. Table \ref{Tab1} lists 
the new eclipse mid-times and their $O-C$ residuals calculated using this ephemeris. The TESS observations are plotted 
as circles in Figure \ref{Fig1}, with the top and middle panels showing the magnitude as a function of BJD and orbital phase, 
respectively. In this figure, the primary and secondary minima occur at phases 0.0 and 0.5, respectively, which implies that 
the EB system is in a circular orbit.

\section{Binary Modeling}

The TESS light curve for V421 Peg displays the typical shape for a detached binary, with both minima appearing as partial 
rather than total eclipses. We modeled individual S57 data of the program target with the Wilson-Devinney (W-D) code 
\citep{Wilson+1971, Kallrath2022}, in the same way as the previously-studied detached DEBs showing multiperiodic pulsations 
\citep{Lee+2021a, Lee+2021b}. The binary mass ratio of $q$ = $M_{\rm B}/M_{\rm A}$ = $K_{\rm A}/K_{\rm B}$ is of utmost 
importance when modeling the target star observables, but it is difficult to derive reliably from light curves alone in 
partially-eclipsing detached binaries. Light curve synthesis combined with spectroscopic measurements allows us to directly 
determine the mass and radius for each component star without any assumptions. 

For this synthesis, we used the semi-amplitude velocities ($K_{\rm AB}$), surface temperatures ($T_{\rm eff,A}$), and 
projected rotational rates ($v_{\rm AB}$sin$i$) measured from the echelle spectra by \citet{Ozdarcan+2016}. The mass ratio was 
set to be $q$ = 0.850 $\pm$ 0.017 from $K_{\rm A}$ = 96$\pm$1 km s$^{-1}$ and $K_{\rm B}$ = 113$\pm$2 km s$^{-1}$. 
The effective temperature of V421 Peg A, which is eclipsed at the primary minimum, was kept fixed at their spectral measurement 
$T_{\rm eff,A}$ = 7250$\pm$120 K, and the secondary star temperature was adjusted to optimize the high-precision TESS light curves. 
\citet{Ozdarcan+2016} assumed synchronous rotations with orbital motion, while we used the rotation-to-orbit velocity ratios of 
$F_{\rm A}$ = 1.04$\pm$0.12 and $F_{\rm B}$ = 1.23$\pm$0.16, which is faster than the synchronous rotations of $v_{\rm A,sync}$ 
= $25.7\pm0.3$ km s$^{-1}$ and $v_{\rm B,sync}$ = $22.1\pm0.3$ km s$^{-1}$ computed from our measurements in Table \ref{Tab2}. 
The limb-darkening coefficients ($x$, $y$) were interpolated from the values updated in \citet{van1993} using the logarithmic law. 
The initial values of most of the other parameters were set with reference to the analysis results of \citet{Ozdarcan+2016}. 

\begin{figure}
\begin{center}
\includegraphics[]{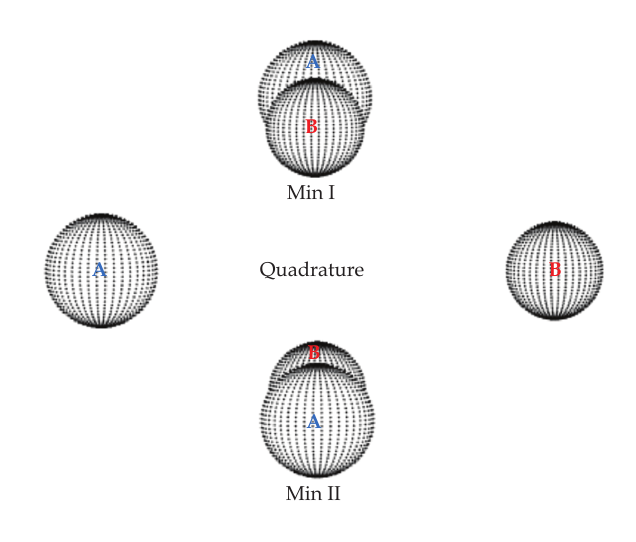}
\end{center}
\caption{Roche-lobe geometrical surfaces of V421 Peg at three orbital phases (from top to bottom, 0.0, 0.25, and 0.50). 
The A and B represent the primary and secondary components, respectively. \\
{Alt text: Geometrical representations of the Roche lobe surfaces.} }
\label{Fig2}
\end{figure}

The binary star modeling was carried out iteratively until the corrections for all free parameters were not greater than 
their 1-$\sigma$ errors. The synthetic curve obtained through this modeling is plotted as a solid line on 
the middle panel of Figure \ref{Fig1}, and the corresponding residuals are presented in the bottom panel. We cannot fully 
understand the residual changes, but they are likely related to stellar pulsations. Table \ref{Tab2} summarizes the model 
parameters optimized for the high-precision TESS data. Our modeling shows that the DEB system exists as a detached configuration 
with both components filling approximately 40 \% of their inner Roche lobes. The Roche-lobe geometrical surfaces of V421 Peg are 
illustrated in Figure \ref{Fig2}. 

The target's absolute properties were evaluated by combining our light curve parameters with the velocity semi-amplitudes  
$K_{\rm A}$ and $K_{\rm B}$ taken from \citet{Ozdarcan+2016}. For this calculation, the Sun's temperature and bolometric magnitude 
were set to $T_{\rm eff}$$_\odot$ = 5780 K and $M_{\rm bol}$$_\odot$ = +4.73, respectively, and the bolometric corrections (BCs) 
were given by the temperature correlation \citep{Torres2010}. Our measurements, summarized at the bottom of Table \ref{Tab2}, 
agree well with those from \citet{Ozdarcan+2016} within an error margin of 1 $\sigma$. Thanks to the high-precision TESS data, 
the radius measurements for both components are accurate to about 1 \%. Incorporating $V$ = +8.290$\pm$0.030 and $E(B-V)$ = 
0.025$\pm$0.018 provided in the TESS v8.2 catalogue \citep{Paegert+2022}, we estimated the geometric distance to V421 Peg to be 
159$\pm$6 pc. This is concurrent with the reciprocal distance of 153.8$\pm$0.6 pc for the Gaia DR3 parallax 6.505$\pm$0.025 mas 
with RUWE=0.964 \citep{Gaia2022}, indicating a good astrometric solution \citep{Stassun+2021}.

\section{Pulsational Characteristics}

\begin{figure}
\begin{center}
\includegraphics[scale=0.9]{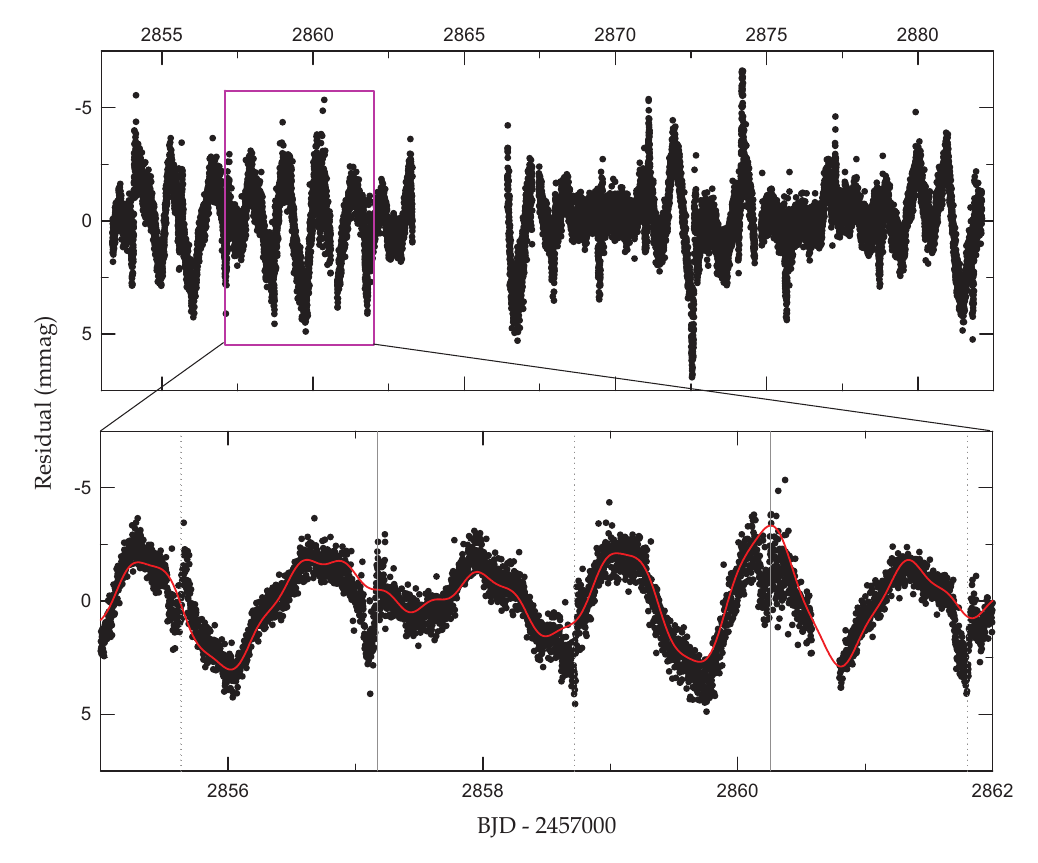}
\end{center}
\caption{Light curve residuals after subtracting the binary effect from the W-D modeling fit to the TESS data. The lower panel 
shows a zoomed-in view of the residuals marked using the inset box in the upper panel. The synthetic curve is computed from 
the 9-frequency fit to the outside-eclipse part of the residuals. The vertical solid and dotted lines indicate the primary 
and secondary minima measured from the observed TESS data, respectively. \\
{Alt text: TESS residual lights distributed in BJD instead of orbital phase.} }
\label{Fig3}
\end{figure}

Figure \ref{Fig3} shows the light residuals from the W-D model fit distributed in BJD, where we can see cycle-to-cycle pulsations 
with a full amplitude of $\sim$5 mmag. The PERIOD04 software \citep{Lenz+2005} was applied to the non-eclipsing part (phases 0.036$-$0.464 
and 0.536$-$0.964) of these residuals for multi-frequency detection up to the Nyquist limit of 360 day$^{-1}$. The pre-whitening 
sequence was repeated until no further meaningful signals were found \citep{Lee+2014}. As a consequence, we extracted a total of 
nine frequencies with amplitudes approximately five times greater than the noise levels computed over a range of 5 day$^{-1}$ 
around each signal in the Fourier spectra, as suggested by the \citet{Baran+2021} simulations. The synthetic curve for 
the extracted multiple frequencies appears as a solid line in the lower panel of Figure \ref{Fig3}. 

\begin{figure}
\begin{center}
\includegraphics[]{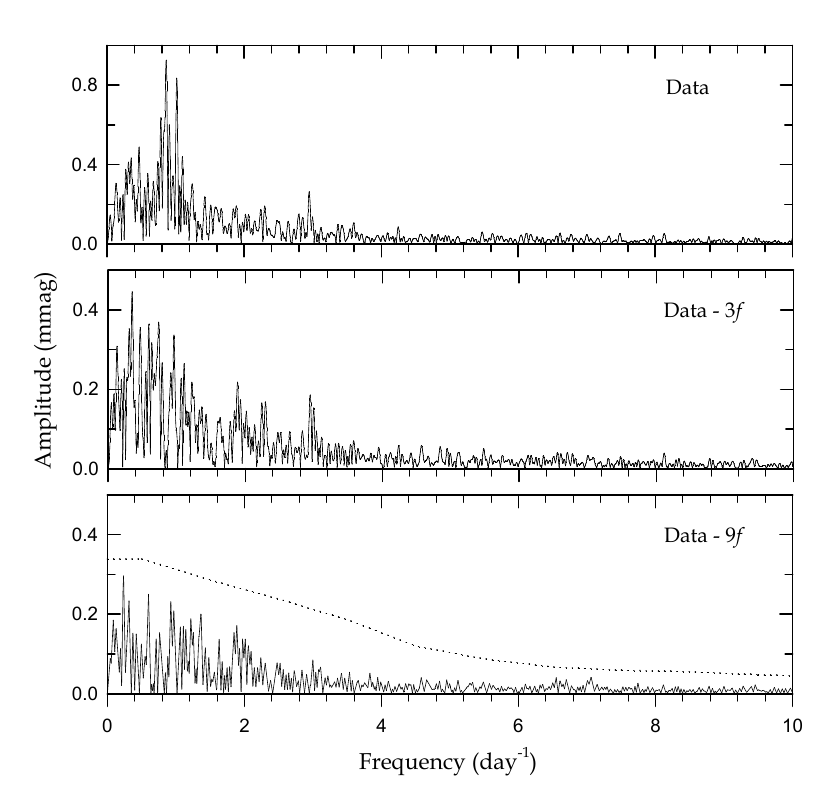}
\end{center}
\caption{Amplitude spectra before (top panel) and after prewhitening the first three frequencies (middle) and all nine frequencies 
(bottom) from the PERIOD04 program for the out-of-eclipse residuals. The dotted line in the bottom panel corresponds to five times 
the noise spectrum. \\
{Alt text: PERIOD04 periodogram for the residual light curve of V421 Peg.} }
\label{Fig4}
\end{figure}

The multi-frequency solution for V421 Peg is given in Table \ref{Tab3}, and the amplitude spectra from the PERIOD04 periodogram 
are displayed in Figure \ref{Fig4}. The middle and bottom panels show the pre-whitening of the first three frequencies, followed 
by whitening of all nine frequencies. We did not detect any conspicuous pulsations in the frequency range higher than 3 day$^{-1}$, 
indicating that $\delta$ Sct-like pulsations are not present in V421 Peg \citep{Grigahcene+2010,Uytterhoeven+2011}. 
The Rayleigh criterion for the 28.8-day S57 dataspan was used as a frequency resolution to distinguish aliasing signals in 
the extracted frequencies. The results of this process are summarized in the last column of Table \ref{Tab3}. The frequencies 
$f_4$ and $f_8$ appear as the orbital frequency ($f_{\rm orb}$ = 0.3239 day$^{-1}$) and its doubling, respectively. 
In addition, $f_5$, $f_7$, and $f_9$ can be considered to be aliasing frequencies associated with $f_{\rm orb}$ or $f_4$. 
These five signals may be due to sampling artifacts resulting from the use of the out-of-eclipse residuals or the lack of trend 
correction in the TESS data. V421 Peg is thought to pulsate at four independent frequencies of $f_1$, $f_2$, $f_3$, and $f_6$, 
which are most likely the signs of $\gamma$ Dor pulsations in the A/F-type intermediate-mass MS band. 

To investigate what components are responsible for the multi-period pulsations, we analyzed the light residuals from the primary 
and secondary eclipses separately, but the partial eclipses and the short duration of the in-eclipse data made it difficult to 
determine.  In the H-R diagram displaying $\gamma$ Dor components in EBs \citep{Ozdarcan+2016,Lee2016}, the primary component of 
V421 Peg is located in the region where the $\delta$ Sct and $\gamma$ Dor pulsators coexist, while the secondary companion appears 
to be relatively low-luminosity below the zero-age MS. We believe that the $\gamma$ Dor pulsations discovered in this work could 
originate from both components, but the primary star is more likely to be the main source of the pulsating signals, as suggested 
in \citet{Ozdarcan+2016}. By applying the primary star density to the relation $Q_i$ = $f_i$$\sqrt{\rho_{\rm A} / \rho_\odot}$, 
we computed the pulsation constants for independent frequencies to be $Q_1$ = 0.745 days, $Q_2$ = 0.631 days, $Q_3$ = 0.775 days, 
and $Q_6$ = 0.876 days. The $Q$ values, pulsation frequencies, and position on the H-R diagram show that the multiperiodic signals 
are $\gamma$ Dor variables of V421 Peg A.

\section{Conclusion}

For the DEB system V421 Peg, we conducted an in-depth study of the short-cadence TESS data to update the binary parameters and 
characterize its pulsations. The high-quality light curve, showing partial eclipses, was solved using a synthetic binary model 
combined with the spectroscopic measurements from \citet{Ozdarcan+2016}. The modeling results confirm that the program target 
is a circular-orbit detached system, whose component stars fill $\sim$40 \% of their limiting lobes and have masses of 
1.589$\pm$0.044 $M_\odot$ and 1.350$\pm$0.032 $M_\odot$, radii of 1.571$\pm$0.019 $R_\odot$ and 1.347$\pm$0.019 $R_\odot$, and 
luminosities of 6.11$\pm$0.43 $L_\odot$ and 3.85$\pm$0.27 $L_\odot$. These fundamental parameters are consistent with those of 
\citet{Ozdarcan+2016} within their error range, but our analysis of space-based TESS data allowed us to more precisely measure 
the radius of each component. The predicted parallax of 6.29 $\pm$ 0.24 mas from our EB distance (159$\pm$6 pc) is in good 
agreement with the Gaia trigonometric measurement of 6.505$\pm$0.025 mas. Further refinements to the binary properties can be 
achieved by a new high-resolution spectroscopic study, which would provide more double-lined RVs distributed across 
the orbital phases and allow better measurements of atmosphere parameters. 

The pulsation features are clearly visible in the TESS residual lights after subtracting the binary star model. We performed 
multi-frequency analysis on the non-eclipse residuals and extracted nine significant signals in the frequency range lower 
than 3 day$^{-1}$, which fall into the $\gamma$ Dor pulsation domain \citep{Uytterhoeven+2011}. Based on the Rayleigh criterion, 
we excluded possible aliasing frequencies from the extracted frequencies and identified the remaining four ($f_1$, $f_2$, $f_3$, 
and $f_6$) as independent pulsations. The positions of V421 Peg A and B on the H-R diagram suggest that the primary star is 
the main source of these pulsations. The pulsating frequencies of 0.732$-$1.015 day$^{-1}$ and the $Q$ values of 0.631$-$0.876 days 
are strongly reminiscent of the gravity modes of $\gamma$ Dor stars \citep{Grigahcene+2010,Uytterhoeven+2011,Antoci+2019}. 
These results demonstrate that the light changes of V421 Peg present in the TESS residuals are produced by multiperiodic pulsations, 
and that the primary component is a $\gamma$ Dor-type pulsator with super-synchronous rotation in the MS phase. 
Unlike semi-detached binaries, which can be driven by mass transfer, the super-synchronous rotation is uncommon and 
poorly understood in detached MS binaries with both short orbital periods and circular orbits, such as 421 Peg \citep{Lurie+2017}. 
This fast rotation could be a result of primordial star formation or differential rotation driven by changes in the internal 
structure of the EB components \citep{Koenigsberger+2021,Britavskiy+2024}. 
This study is a continuation of attempts to look for and characterize pulsation signals in DEBs. Such samples are not only promising 
for validating theoretical stellar models, but also for understanding the binary effect on pulsations. 


\begin{ack}
This paper has made use of the ultra-precise TESS public archives and the Simbad database maintained at CDS, Strasbourg, France. 
We gratefully acknowledge the support by the KASI grant 2025-1-830-05. 
\end{ack}

\onecolumn{}
\clearpage

\clearpage 
\begin{table}
\tbl{TESS Eclipse Timings for V421 Peg. }{%
\begin{tabular}{lcccc}
\hline
BJD               & Error           & $E$           & $O$--$C$           & Min  \\ 
\hline
2,459,854.082393  & $\pm$0.000018   & 0.0           & $-$0.000125        & I    \\
2,459,855.626412  & $\pm$0.000021   & 0.5           & $+$0.000116        & II   \\
2,459,857.170116  & $\pm$0.000012   & 1.0           & $+$0.000041        & I    \\
2,459,858.713897  & $\pm$0.000017   & 1.5           & $+$0.000044        & II   \\
2,459,860.257819  & $\pm$0.000017   & 2.0           & $+$0.000187        & I    \\
2,459,861.801504  & $\pm$0.000015   & 2.5           & $+$0.000094        & II   \\
2,459,866.432918  & $\pm$0.000075   & 4.0           & $+$0.000173        & I    \\
2,459,867.976572  & $\pm$0.000010   & 4.5           & $+$0.000048        & II   \\
2,459,869.520309  & $\pm$0.000011   & 5.0           & $+$0.000007        & I    \\
2,459,871.063950  & $\pm$0.000013   & 5.5           & $-$0.000131        & II   \\
2,459,872.607610  & $\pm$0.000009   & 6.0           & $-$0.000249        & I    \\
2,459,874.151335  & $\pm$0.000013   & 6.5           & $-$0.000303        & II   \\
2,459,875.695446  & $\pm$0.000012   & 7.0           & $+$0.000030        & I    \\
2,459,877.239315  & $\pm$0.000015   & 7.5           & $+$0.000120        & II   \\
2,459,878.783040  & $\pm$0.000009   & 8.0           & $+$0.000067        & I    \\
2,459,880.326991  & $\pm$0.000012   & 8.5           & $+$0.000239        & II   \\
2,459,881.870529  & $\pm$0.000014   & 9.0           & $-$0.000001        & I    \\
\hline
\end{tabular}}\label{Tab1}
\end{table}

\begin{table}
\tbl{Binary Parameters of V421 Peg. }{%
\begin{tabular}{lcc}
\hline
Parameter                         & Primary (A)           & Secondary (B)               \\ 
\hline
$T_0$ (BJD)                       & \multicolumn{2}{c}{2,459,854.082781$\pm$0.000055}   \\   
$P_{\rm orb}$ (day)               & \multicolumn{2}{c}{3.087555$\pm$0.000010}           \\   
$q$                               & \multicolumn{2}{c}{0.850$\pm$0.017}                 \\   
$i$ (deg)                         & \multicolumn{2}{c}{86.244$\pm$0.022}                \\   
$T_{\rm eff}$ (K)                 & 7250$\pm$120          & 6977$\pm$110                \\   
$\Omega$                          & 8.989$\pm$0.023       & 9.136$\pm$0.030             \\   
$\Omega_{\rm in}$$\rm ^a$         & \multicolumn{2}{c}{3.501}                           \\   
$F$                               & 1.04$\pm$0.12         & 1.23$\pm$0.16               \\   
$x$, $y$                          & 0.516, 0.283          & 0.529, 0.281                \\   
$l/(l_{\rm A}+l_{\rm B})$         & 0.6014$\pm$0.0022     & 0.3986                      \\   
$r$ (pole)                        & 0.1228$\pm$0.0007     & 0.1053$\pm$0.0010           \\   
$r$ (point)                       & 0.1233$\pm$0.0007     & 0.1058$\pm$0.0010           \\   
$r$ (side)                        & 0.1230$\pm$0.0007     & 0.1055$\pm$0.0010           \\   
$r$ (back)                        & 0.1233$\pm$0.0007     & 0.1057$\pm$0.0010           \\   
$r$ (volume)$\rm ^b$              & 0.1230$\pm$0.0007     & 0.1055$\pm$0.0010           \\ [1.0mm]
\multicolumn{3}{l}{Absolute parameters:}                                                \\   
$M$ ($M_\odot$)                   & 1.589$\pm$0.044       & 1.350$\pm$0.032             \\   
$R$ ($R_\odot$)                   & 1.571$\pm$0.019       & 1.347$\pm$0.019             \\   
$\log$ $g$ (cgs)                  & 4.247$\pm$0.016       & 4.310$\pm$0.016             \\   
$\rho$ ($\rho_\odot$)             & 0.411$\pm$0.019       & 0.553$\pm$0.027             \\   
$L$ ($L_\odot$)                   & 6.11$\pm$0.43         & 3.85$\pm$0.27               \\   
$M_{\rm bol}$ (mag)               & 2.77$\pm$0.08         & 3.27$\pm$0.08               \\   
BC (mag)                          & 0.04$\pm$0.01         & 0.03$\pm$0.01               \\   
$M_{\rm V}$ (mag)                 & 2.73$\pm$0.08         & 3.24$\pm$0.08               \\   
Distance (pc)                     & \multicolumn{2}{c}{159$\pm$6}                       \\   
\hline
\end{tabular}}\label{Tab2}
\begin{tabnote}
\footnotemark[a]Potential for the inner critical Roche surface. 
\footnotemark[b]Mean volume radius. 
\end{tabnote}
\end{table}

\begin{table}
\tbl{Results of the multiple frequency analysis for V421 Peg$\rm ^{a,b}$. }{%
\begin{tabular}{lcccrc}
\hline
             & Frequency              & Amplitude           & Phase           & SNR$\rm ^c$    & Remark              \\ 
             & (day$^{-1}$)           & (mmag)              & (rad)           &                &                     \\
\hline                                                                                         
$f_{1}$      &  0.8606$\pm$0.0006     & 1.04$\pm$0.09       & 3.44$\pm$0.26   & 16.20          &                     \\
$f_{2}$      &  1.0154$\pm$0.0007     & 0.87$\pm$0.08       & 0.80$\pm$0.29   & 13.62          &                     \\
$f_{3}$      &  0.8276$\pm$0.0011     & 0.61$\pm$0.10       & 6.04$\pm$0.47   &  9.40          &                     \\
$f_{4}$      &  0.3477$\pm$0.0013     & 0.49$\pm$0.10       & 6.28$\pm$0.57   &  7.09          & $f_{\rm orb}$       \\
$f_{5}$      &  0.4660$\pm$0.0020     & 0.34$\pm$0.10       & 0.64$\pm$0.84   &  5.03          & $f_3-f_4$           \\
$f_{6}$      &  0.7320$\pm$0.0010     & 0.62$\pm$0.09       & 5.02$\pm$0.44   &  9.51          &                     \\
$f_{7}$      &  0.6798$\pm$0.0013     & 0.49$\pm$0.09       & 5.36$\pm$0.55   &  7.49          & 2$f_4$              \\
$f_{8}$      &  0.6346$\pm$0.0016     & 0.42$\pm$0.10       & 2.08$\pm$0.68   &  6.33          & 2$f_{\rm orb}$      \\
$f_{9}$      &  2.9453$\pm$0.0024     & 0.21$\pm$0.07       & 2.90$\pm$1.01   &  4.98          & 4$f_7$              \\
\hline
\end{tabular}}\label{Tab3}
\begin{tabnote}
\footnotemark[a]Frequencies are listed in order of detection. \\
\footnotemark[b]Parameters' errors were estimated following \citet{Kallinger+2008}. \\
\footnotemark[c]Signal to noise amplitude ratios calculated in a range of 5 day$^{-1}$ around each frequency. \\
\end{tabnote}
\end{table}

\end{document}